# MEASURES OF RADIOACTIVITY: A TOOL FOR UNDERSTANDING STATISTICAL DATA ANALYSIS

**Vera Montalbano,** *Phys. Department, University of Siena*
**Sonia Quattrini,** *Istituto Tecnico Tecnologico Sarrocchi, Siena*

**Abstract**
A learning path on radioactivity in the last class of high school is presented. An introduction to radioactivity and nuclear phenomenology is followed by measurements of natural radioactivity. Background and weak sources are monitored for days or weeks. The data are analyzed in order to understand the importance of statistical analysis in modern physics.

## 1. An early experience

Since 2006, the Pigelleto's Summer School of Physics is an important appointment for orienting students toward physics in our territorial area (Benedetti 2012).

Forty students from high school are selected to attend a full immersion summer school of physics in the Pigelleto Natural Reserve, on the south east side of Mount Amiata in the province of Siena.

During the 2009 edition, titled *The Achievements of Modern Physics* (Benedetti 2011), a learning path on radioactivity was proposed by one of us (VM) to small groups of students. After a brief introduction to the nuclear phenomenology, they were involved in measures of radioactivity from a weak source of Uranium in order to characterize the emitted ionizing radiation by using an educational device provided by a school.

A teacher (SQ) was impressed by student's involvement and suggested of elaborating a learning path on Nuclear Physics in which the students are active and perform directly measures of radioactivity, in order to propose this activity in her school.

## 2. A learning path on nuclear phenomena

We planned a path which was proposed as a laboratory within the National Plan for Science Degree and the school obtained founds for it from a regional project supported by Tuscany (Progetto Ponte, i.e. Bridge Project) whose purpose was to promote relations between Universities and High Schools for orienting students.

In order to integrate this learning path in the ordinary school program, the activities were planned for the last year of high school (5a Liceo Tecnologico) after the lessons on electromagnetism.

Teachers make few lessons in class in which presented and discussed the phenomenology of nuclear physics. Afterward, the experimental setup was presented in order to perform measure of background radioactivity in the laboratory.

The next step was to divide students in small groups that are involved in measures of alpha, beta and gamma emission from weak sources of uranium and a very weak source with trace of uranium ore.

During measurements, it rose soon the necessity of performing many measures and the problem of valuating uncertainties. The students were ready to recall all their knowledge on statistics for facing this problem. Tab. 1 shows the topics treated by the teacher in class, in physics laboratory and in computer lab.

### *2.1. A brief introduction to nuclear phenomena*

The lessons give an historical sight on the discovery of radioactivity, and connect the previous knowledge of students to these phenomena so far from their daily experience.

Many examples and easy computations are performed in order to clarify concepts such as binding force, the relative intensities of fundamental interactions at atomic and nuclear distances, mass defect and binding energy, equivalence mass-energy and so on.

Topics related to fission and fusion can be very stimulating for students, specially for our society in these times of discussion about use of nuclear energy and security.

*Table 1: Covered topics*

| *A brief introduction to nuclear phenomena* | *Physics Laboratory* | *An introduction to statistical data analysis* | *Computer Lab* |
| --- | --- | --- | --- |
| Discovery of radioactivity | Background measures | Distributions: Bernoulli, Poisson, Gauss | Verify that radioactive decay follows Poisson's distributions |
| Pierre e Marie Curie | Measurement of Uranium sources at different distances | Representation of casual phenomena | Testing samples by giving 10, 20, 30, 50 data from the background |
| Neutron discovery and atom structure | Measurement of Uranium sources at different distances | Mean, median, variance and standard deviation | Compare means and standard deviations of previous samples with the mean and standard deviation of very long measure (more then 1000 data) |
| Dimensions of atom and its constituents | $\alpha$, $\beta$ and $\gamma$ rays measurement | Bernoulli, Poisson and Gauss distributions | Verify that for mean values greater than … the Poisson's distribution becomes indistinguishable from the normal one |
| Natural radioactivity phenomenology (in particular $\alpha$, $\beta$ and $\gamma$ emission) | $\alpha$, $\beta$ and $\gamma$ rays measurement with different shields | Examples | Compare measures of background and very weak source |
| Isotopes | Measures with a small Uranium source | The existence of a parent distribution | |
| Forces into atoms and nuclei (how strong or weak with numerical examples) | Measures with a smaller Uranium source | What is a good statistical sample | |
| Nuclei can die… | Measures with a very weak Uranium ore source | Mean and variance of a statistical sample | |
| Law of radioactive decay | | | |
| Natural radioactive chains | | | |
| Binding energy and mass of a bound state | | | |
| Equivalence mass-energy | | | |
| Binding energy per nucleon | | | |
| Fission and related topics | | | |
| Fusion and related topics | | | |

## 2.2. The experimental setup

The measures are realized by using a commercial Geiger counter usually used for dosimetry. It can detect alpha, beta and gamma rays and it is possible to collect data without any unit conversion (counts) and download them into a computer.

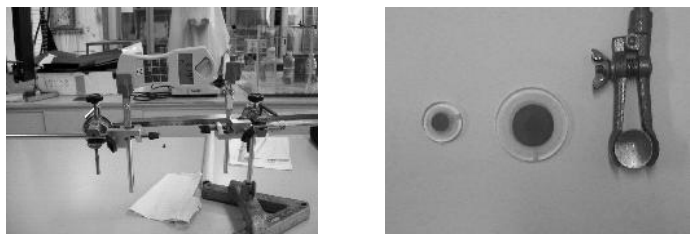

*Figure 1: a) the experimental setup b) radioactive source in their supports*

The detector and the sources are aligned by means of supports on an optical bench as shown in Fig. 1.a. The sources are shown with their supports in Fig. 1.b: (from left) small Uranium sheets in plastic holder ($m_1$ = 0.152 g), other Uranium sheets ($m_2$ = 0.888 g), fluorescent marble.

### 2.3. An introduction to statistical data analysis

We started from a question. There are limits to the precision of a measurement? By answering to this question, we can fix the cases in which the use of statistical data analysis make sense.

The first step is to recall all previous knowledge of students on statistics (there are many but almost never used in physics by teachers), such as mean, variance, standard deviation, distributions (Bevington 1969). At this point, it is possible to introduce the pupils to some elements of sampling theory[1]. They are ready to deal with the statistical data analysis of our measures.

## 3. Measures and statistical data analysis

The data are collected in the meantime by the Geiger detector and can be downloaded in a computer. The teacher can divide the data in sets in order of giving different tasks to any group of students in computer lab.

In Fig. 2, a confront between two slightly different measures can be made. Fig. 1.a shows the data which are divided into to groups ( $N_1$ = 188, with mean 182, standard deviation 16, sample standard deviation 1 and $N_2$ = 437, with mean 194.5, standard deviation 14, sample standard deviation 0.7). It is easy to compute the difference between means but it is overwhelmed by the uncertainty if one uses standard deviation, on the contrary the use of sample standard deviation allows to estimate quantitatively the difference.

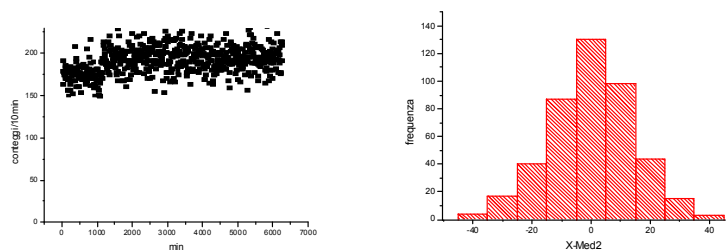

*Figure 2: a) the data from Geiger detector, b) frequency histogram for $N_2$ data*

## 4. Conclusions

This learning path on nuclear phenomena was performed on three last classes in 2010 and 2011. We presented the revised path, after analysing the learning difficulties encountered by students in these years. All classes made the introduction to nuclear phenomena in class and in laboratory, only one class made some activity in computer lab. In our opinion, this path is very interesting for students but hardly feasible in last class due to graduation exam. It can be an excellent optional proposal for interested students but if a teacher want to implement it in class it needs to be revised the presentation of statistical topics in previous years.

---

[1] Stat Trek 2011 for a synthetic review